# EFFECT OF NODE MOBILITY ON AOMDV PROTOCOL IN MANET

Indrani Das[1], D.K Lobiyal[2] and C.P Katti[3]

[1,2,3]School of Computer and Systems Sciences
Jawaharlal Nehru University, New Delhi, India.

## ABSRACT

*In this paper, we have analyzed the effect of node mobility on the performance of AOMDV multipath routing protocol. This routing protocol in ad hoc network has been analyzed with random way point mobility model only. This is not sufficient to evaluate the behavior of a routing protocol. Therefore, in this paper, we have considered Random waypoint, Random Direction and Probabilistic Random Walk mobility Model for performance analysis of AOMDV protocol. The result reveals that packet delivery ratio decreases with the increasing node mobility for all mobility models. Also, average end-to-end delay is also vary with varying node speed, initially upto 20 nodes in all mobility models delay is minimum.*

## 1. INTRODUCTION

A Mobile Ad-Hoc Network (MANET) is a network where more than two autonomous mobile hosts (mobile devices i.e. mobile phone, laptop, iPod, PDAs etc.) can communicate to each other without any mean of fixed infrastructure. When source($S$) node want to send some data toward the destination ($D$), if they are fall in the same transmission range only can directly communicate with each other. Otherwise with the help of intermediate nodes communication can be established. Any node may join and leave the network in any point of time, therefore the topology of the network changes frequently. In this network some scarce resources like battery power of mobile devices, bandwidth of network. The depletion of battery power may affect lifetime of the whole network as well as individual node existence in the network. Due to dynamic topology and other network constraint routing in MANET is a challenging issue. Single path routing is not always sufficient to disseminate data to the destination. Therefore; multipath routing comes into existence to overcome the problem of single path routing.

In this paper we have considered various mobility models for proper and in depth analysis of AOMDV protocol. In literature we have discussed various works related to AOMDV protocol and brief about various multipath routing protocols. Most of the work carried out based on random waypoint mobility model. Therefore, we have analyzed AOMDV protocol with various network parameters and mobility models. Finally, we have computed packet delivery ratio and average end-to-end delay with varying node speed for individual mobility models.

The rest of the paper is organized as follows. In section II we have discussed various works related to multipath routing. In section III, various mobility models and AOMDV routing protocol-briefly discussed. Results analysis and simulation work is presented in Section IV and finally, we have concluded the paper in Section V.

## 2. RELATED WORKS

Multipath routing overcomes various problems occurs while data delivered through a single path. The multipath routing protocols are broadly classified based on on-demand, table driven, and hybrid. The following multipath routing protocols are used in MANETs. In [1] authors have compared the performance of AOMDV and OLSR routing protocol with Levy-Walk and Gauss-Markov Mobility Model. For the analysis they have considered varying mobility speed and the traffic load in the network. Their results show that AOMDV protocol achieved higher packet delivery ratio and throughput compared to OLSR. Further, OLSR has less delay and routing overhead at varying node density. In [2] authors only compared AOMDV and AODV routing protocol with random way point mobility model. Different traffic source like TCP and CBR is considered. The result shows that with increasing traffic both routing protocols performance degraded. In M-DSR (Multipath Dynamic Source Routing) [5, 21] is an on demand routing protocol based on DSR [12]actually it  is a multipath extension of DSR. In SMR (Split Multipath Routing) [5, 15] is an on demand routing protocol and extension of well- known DSR protocol.  The main aim of this protocol is to split the traffic into multiple paths so that bandwidth utilization goes in an efficient manner. In GMR (Graph based Multipath Routing) [5, 9] protocol based on DSR, a destination node compute disjoint path in the network using network topology graph.In MP-DSR [5, 13, 16] is based on DSR; it is design to improve QoS support with respect to end-to-end delay. In [10,19] authors have proposed an on-demand multipath routing protocol AODV-BR. But to establish multipath it does not spend extra control message. This protocol utilizes mesh structure to provide multiple alternate paths. In [8] authors have considered node-disjoint and link-disjoint multi-path routing protocol for their analysis. The various mobility models Random Waypoint, Random Direction, Gauss-Markov, City Section and Manhattanmodels are considered. Through the thorough analysis they have shown that in Gauss markov mobility model multipath formation is less but path stability is high. (The random direction model form larger number of multipath.) In [14] authors have considered AODV and AOMDV protocol for their performance analysis with random waypoint model.  The result shows that AOMDV has more routing overhead and average end to end delay compared to AODV. But AOMDV perform better in term of packets drops and packet delivery. In [17] various energy models with Random Waypoint Mobility Model,Steady State mobility model is used to analyze the energy overhead in AOMDV, TORA and OLSR routing protocols. Results show that TORA protocol has highest energy overhead in all the energy models.In [22] performance of AOMDV protocol is analyzedfor different mobility models to investigate how this protocol behaves in different mobility scenario. The results show that with increasing node density, packet delivery ratio increases but with increasing node mobility packet delivery ratio decreases.

## 3. DESCRIPTION OF ROUTING PROTOCOL AND MOBILITY MODELS

In this section we have discussed brief about AOMDV routing protocol and various mobility models considered for simulation work.

### 3.1 Ad Hoc On Demand Multipath Distance Vector (AOMDV)

Ad Hoc On Demand Multipath Distance Vector (AOMDV) [3, 5, 6, 11] protocol is a multipath variation of AODV protocol. The main objective is to achieve efficient fault tolerance i.e. quickly recovery from route failure. The protocol computes multiple link disjoint loop free paths per route discovery. If one path fails the protocol choose alternate route from other available paths. The route discovery process is initiated only when to a particular destination fails. When a source needs a route to destination will floods the RREQ for the destination and at the intermediate nodes all duplicate RREQ are examined and each RREQ packet define an alternate route. However, only link disjoint routes are selected (node disjoint routes are also link disjoint). The destination node replies only k copies of out of many link disjoint path, i.e. RREQ packets arrive through unique neighbors, apart from the first hop are replied. Further, to avoid loop 'advertised hop count' is used in the routing table of node .The protocol only accepts alternate route with hop count less than the advertised hop count. A node can receive a routing update via a RREQ or RREP packet either forming or updating a forward or reverse path .Such routing updates received via RREQ and RREP as routing advertisement.

## 3.2 Mobility Models

Mobility pattern of node plays a vital role in evaluation of any routing protocol in MANET. We have considered various categories mobility models for acceptability of routing protocol. The following mobility model we have considered in simulation work.

### 3.2.1 Random Waypoint Model

The Random Waypoint (RWP) mobility model [4,7] is the only model which is used in maximum cases for evaluation of MANET routing protocols. In this model nodes movement depends on mobility speed, and pause time. Nodes are moving in a plane and choose a new destination according to their speed. Pause time indicate that a node to wait in a position before moved to new position.

### 3.2.2 Probabilistic Random Walk Model

In this model [4,7]nodes next position is determined by set of probabilities. A node can be move forward, backward or remain in x and y direction depends on the probability defined in probability matrix. There are three state of node is defined by 0 (current position), 1 (previous position) and 2 (next position). Where, in the matrix P (a,b) means the probability that an node will move from state a to state b.

### 3.2.3 Random Direction Model

The random direction model [4,7] is the further modification of Random waypoint mobility model.This model overcome the density wave problem occur in random waypoint model, where clustering of nodes occur in a particular area of simulation. In Random Waypoint model this density occurs in the center of the simulation area. Here, nodes are move upto the boundary of the simulation area before moving to a new location with new speed and direction. When nodes are reached to the boundary of simulation area, before changing to new position it pauses there for sometimes. The random direction it chooses from 0 to 180 degrees. The same process is continued till the simulation time.

## 4. SIMULATION SETUP AND RESULT ANALYSIS

For the simulation works we have used Bonn-Motion mobility generator [18] to generate the mobility of nodes based on various mobility models. The most popular network simulator NS-2.34 [20] has beenused for simulation. Finally, in table-1 and table-2 different simulation parameters and their values have been shown respectively..

Table 1. Simulation Parameters

| Parameter | Specifications |
|---|---|
| MAC Protocol | IEEE 802.11 |
| Routing Protocol | AOMDV |
| Radio Propagation Model | Two-ray ground reflection model |
| Channel type | Wireless channel |
| Antenna model | Omni-directional |
| Mobility Models | Random Waypoint, Random Direction, Probabilistic Random Walk |

Table 2. Values of Simulation Parameters

| Parameter | Values |
|---|---|
| Simulation Time | 1000s |
| Simulation Area  (X *Y ) | 1000 m x1000 m |
| Transmission Range | 250 m |
| Bandwidth | 2 Mbps |
| No. of Nodes | 10,20,30,40,50,60,70,80.90, 100 |
| Node speed | 10,20,30,40 m/s |

Figure1 shows Packet delivery ratio with Random Waypoint Mobility Model for different node speeds. In this model AOMDV gives better packet delivery ratio with increasing node density. But with the increasing node mobility PDR decreases due to frequent link breakage among nodes. The maximum achievable value of PDR is 77.8%.

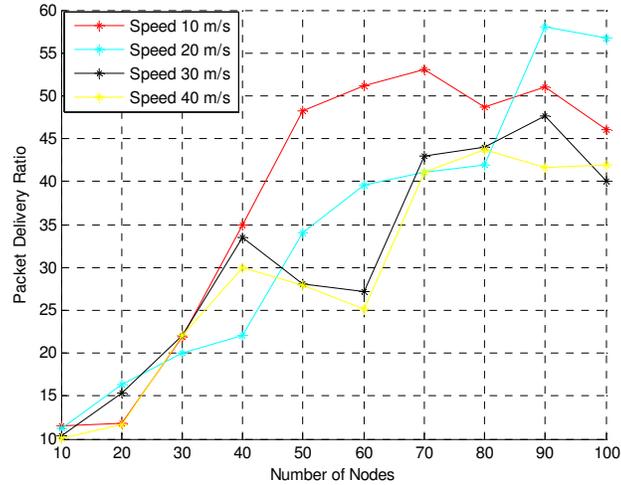

Figure1.Packet delivery ratio with variousnode speed (Random Waypoint Mobility Model).

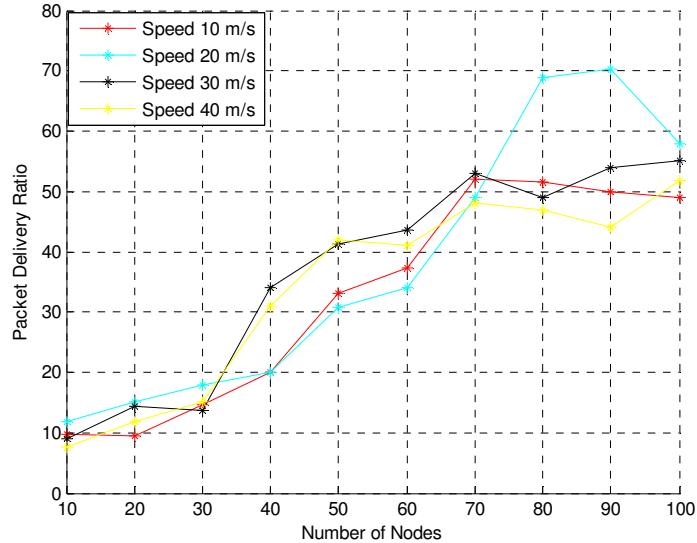

Figure2. Packet delivery ratio with variable node speed(Random Direction Mobility Model).

Figure.2 shows the packet delivery ratio for differentnode mobility for Random Direction Mobili-ty model. In this mobility model PDR decreases with the increasing node mobility. For 90 nodes and speed of 20 m/s maximum value of PDR i.e. 70% is achieved. There is a sudden drop in the PDR as the number of nodes increases beyond 90 due to congestion, sudden link failure etc.At node speed 10 and 20 m/s protocol performance is steady and with increasing speed, the value of PDR reduces.

Figure.3 shows the packet delivery ratio for different node mobility using Probabilistic Random walk mobility model. In this modelAOMDV gives poor performance in term of PDR when node speed increases due to frequent link breakage among nodes. The maximum achievable value of PDR is 80% approximatelyat thenode speed of 10 m/s.

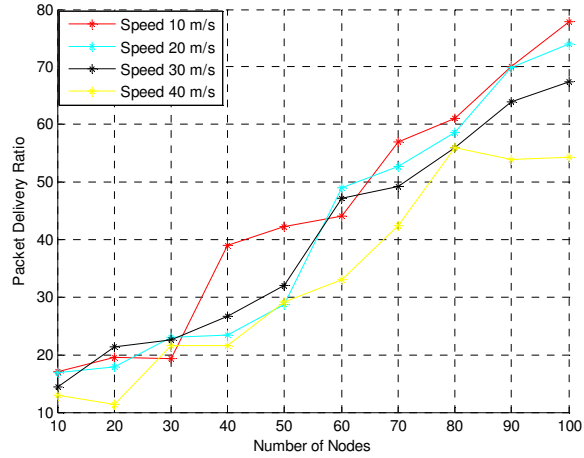

Figure3. Packet delivery ratio with variable node speed (Probabilistic Random Walk Model).

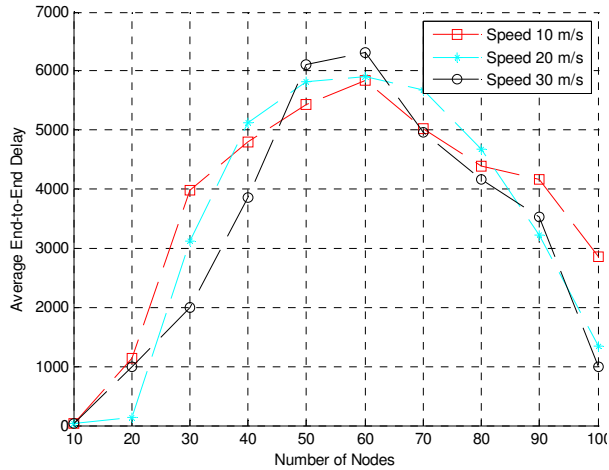

Figure 4.Average end-to-end delay with variable node speed (Random Waypoint Mobility Model).

Figure.4 shows the Average end-to-end delay with variable node speed in random waypoint mobility model. Here, upto 20 nodes in different speed delay is minimum but increases gradually as number of node increases. The maximum delay noticed when speed is 30 m/s and number of node 60. The delay is gradually decreases when number of nodes more than 60 onwards.

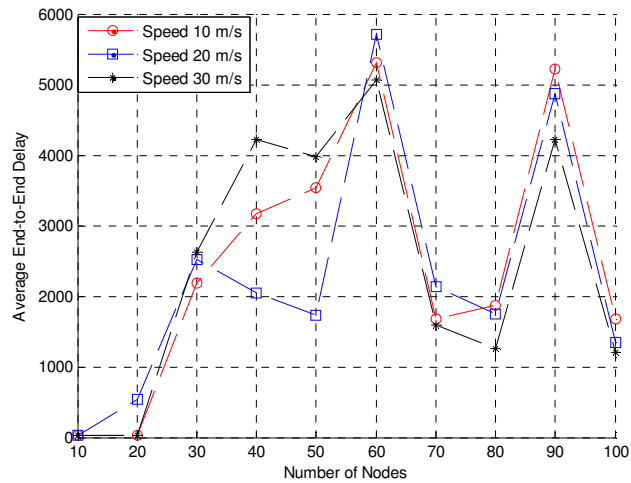

Figure 5.Average end-to-end delay with variable node speed (Random Direction Mobility Model).

Figure.5 shows the Average end-to-end delay with variable node speed in random direction mobility model.Here, upto node 20 in different speed delay is minimum but increases gradually as node increases. The maximum delay noticed when speed is 20 m/s and number of node 60. Further, sudden fall in delay is noticed at number of node is 70 and 80.

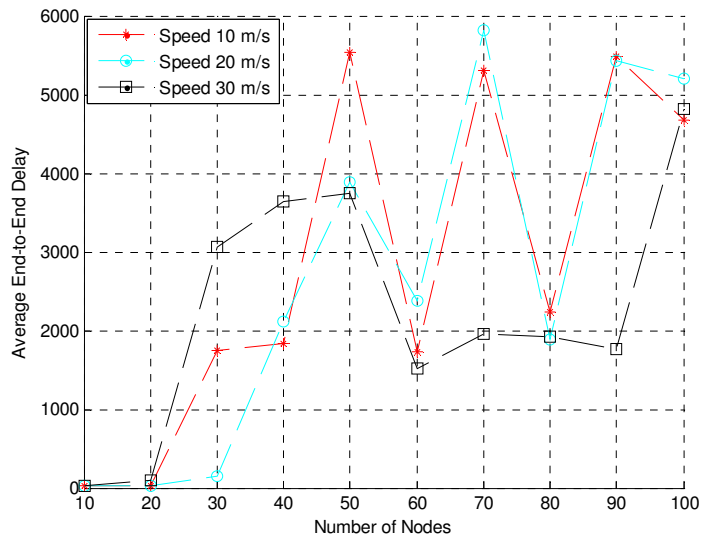

Figure 6.Average end-to-end delay with variable node speed (Probabilistic Random Walk Model).

Figure.6 shows the Average end-to-end delay in probabilistic random walk model with variable node speed. Here, upto 20 nodes in different node speed delay is minimum. As the speed and node vary delay gradually increases. At node speed 30 m/s from node 60 to 90 a consistent delay is noticed.

The overall analysis shows that with high node mobility the value of PDR decreasesfor all mobility models. The average end-to-end delay is gradually increases with increasing speed and nodes, but in random direction and random waypoint mobility models maximum delay noticedat 60 nodes. The delay is gradually decreases 60 nodes onwards. In probabilistic random walk model at node speed 30 m/s average delay in minimum as compare other speed.The overall performance of AOMDV protocol performs better for Randomwaypoint mobility model as compared to other mobility models.

## 5. CONCLUSIONS

We have evaluated the effect of node mobility on the performance of AOMDV multipath routing protocol with different mobility models. For the performance analysis of the protocol packet delivery ratio is computed. It is evident from the results that AOMDV protocol perform better in term of PDR and average end-to-end delay for Random Waypoint mobility model. But it is also noticed that with higher node mobility PDR of AOMDV protocol decreases. In Probabilistic Random walk model upto 70 nodes with various node mobility protocol performs better as compared to others. In future, this multipath protocol can be investigated for different network topologies.

**Authors**

**Indrani Das** did her B. E. and M.Tech in Computer Science. She is working as Assistant Professor in Computer Science department in Assam University (A Central University), Assam, India. Presently, she is perusing her Ph.D from School of Computer and Systems Sciences, Jawaharlal Nehru University, New Delhi, India. Her current research interest includes Mobile Ad hoc Networks and Vehicular Ad hoc Networks.

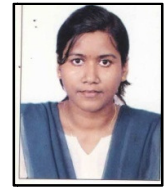

**Daya K. Lobiyal** Received his Bachelor of Technology in Computer Science from Lucknow University, India, in 1988 and his Master of Technology and Ph.D both in Computer Science from Jawaharlal Nehru University, New Delhi, India, 1991 and 1996, respectively. Presently, he is an Associate Professor in the School of Computer and Systems Sciences, Jawaharlal Nehru University, India. His areas of research interest are Mobile Ad hoc Networks, Vehicular Ad Hoc Networks, Wireless Sensor Network and Video on Demand.

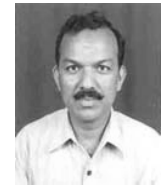

**C. P. Katti** is a professor of computer science at Jawaharlal Nehru University. He received his Ph.D from IIT Delhi. He has published over 30 papers in journals of international repute. His area of research includes parallel computing, ad hoc networks and numerical analysis.

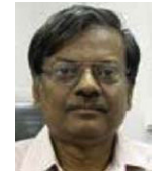